\documentclass{elsart}

\usepackage{epsf} \usepackage{epsfig} \usepackage{latexsym}

\begin{document}

\begin{frontmatter}

  \title{Phase diagram and edge effects in the ASEP with bottlenecks}

\author[koeln]{Philip Greulich}%
\ead{pg@thp.uni-koeln.de} \and
\author[koeln,izks]{Andreas Schadschneider}%
\ead{as@thp.uni-koeln.de}

\address[koeln]{%
  Institut f\"ur Theoretische Physik, Universit\"at zu K\"oln, 
  50937 K\"oln, Germany
}%
\address[izks]{%
  Interdisziplin\"ares Zentrum f\"ur komplexe Systeme,
  53117 Bonn, Germany}%

\date{\today}% It is always \today, today,
\begin{abstract}
  We investigate the totally asymmetric simple exclusion process
  (TASEP) in the presence of a bottleneck, i.e.\ a sequence of
  consecutive defect sites with reduced hopping rate.  The influence
  of such a bottleneck on the phase diagram is studied by computer
  simulations and a novel analytical approach.  We find a clear
  dependence of the current and the properties of the phase diagram
  not only on the length of the bottleneck, but also on its position.
  For bottlenecks near the boundaries, this motivates the
  concept of effective boundary rates. Furthermore the inclusion
    of a second, smaller bottleneck far from the first one has no
    influence on the transport capacity. These results will form the
  basis of an effective description of the disordered TASEP and are
  relevant for the modelling of protein synthesis or intracellular
  transport systems where the motion of molecular motors is hindered
  by immobile blocking molecules.

\end{abstract}

\begin{keyword}
  Nonequilibrium physics, stochastic process, driven lattice gas, disorder
  \PACS 05.40.-a, 02.50.Ey, 45.70.Vn
\end{keyword}
\end{frontmatter}

% main text

%%%%%%%%%%%%%%%%%%%%%%%%%%%%%%%%%%%%%%%%%%%%%%%%%%%%%%%%%%%%%%%%%%%%%%%%%%%%%%
%%%%%%%%%%%%%%%%%%%%%%%%%%%%%%%%%%%%%%%%%%%%%%%%%%%%%%%%%%%%%%%%%%%%%%%%%%%%%%
\section{Introduction}
\label{introduction}

Driven diffusive systems play an important role in statistical
physics.  They not only serve as paradigm for non-equilibrium
behaviour \cite{schmzia,derrida,gunterrev}, but also as models for
transport processes like vehicular traffic \cite{css}, granular flow
through narrow pipes \cite{granflow} and biological transport by motor
proteins \cite{frey1,lipowsky,NOSC1,NOSC2}. The simplest of these
models is the totally asymmetric simple exclusion process (TASEP)
which was first introduced to describe protein polymerization in
ribosomes \cite{mcdonald} (for a recent review, see \cite{mpa}).  The
TASEP exhibits some generic properties like \emph{boundary induced
  phase transitions} \cite{krug1} that also occur in more complex
driven systems. It was solved exactly \cite{derridaEHP,schuetz} and
results can be used for qualitative and quantitative approaches to
other systems. The phase diagram consists of three phases (high
density (h), low density (l), and maximum current phase (m)),
depending on whether the current is limited by the particle input,
output or the transport capacity of the bulk, and is generic for a
large class of driven diffusive systems \cite{koloSKS}.

While homogeneous systems are extensively investigated, systems with
inhomogeneous transition rates are not that well understood,
especially for the case of open boundary conditions. It is well-known that 
in nonequilibrium systems even weak disorder can lead to drastic
changes \cite{Stinch02,Barma06}. 
For driven diffusive systems already a single
\emph{defect site}, i.e.\ a site with reduced hopping rate, can have a global
effect on the stationary state, see e.g.\ \cite{lebowitz,lebowitz2}
for the case of the TASEP with periodic boundary conditions. Here it
was shown that in an intermediate density regime $|\rho-1/2|<\rho_c$
the current becomes density-independent, i.e.\ the flow-density
relation (fundamental diagram) exhibits a plateau. In contrast, in the
limits of low and high densities, the current is unchanged by the
presence of the defect.

Here we consider the TASEP with open boundaries in the presence of a
\emph{bottleneck} of length $l$, i.e.\ a sequence of $l$ consecutive
"slow" bonds\footnote{A bottleneck of length $l=1$ corresponds to a single 
{\em defect}.}, where the hopping rate of the particles
is reduced compared to the rest of the system.  We will study the
effect of such bottlenecks on the phase diagram of the TASEP,
focussing on the \emph{maximal current} that can be maintained by the system
through optimization of the input and output rates, i.e.\ its
\emph{transport capacity}.

In open systems, the main effect of the bottleneck (or defects in general) 
is a decrease of the transport capacity \cite{kolomeiski2}.  
This leads to an enlarged maximum current phase compared to
the pure system. In addition the density profiles in the maximum
current phase exhibit phase separation into a high density and low
density regime separated by a shock.

The effect of defects in systems with open boundaries has been studied
earlier by several authors. So far no exact solutions are known and
one has to rely on computer simulations with Monte Carlo (MC) methods
and mean-field type approximations.

Kolomeisky \cite{kolomeiski2} investigated the TASEP in the presence of a 
single defect site ($l=1$) deep in the bulk. Analytical
results are obtained by a mean field approach which neglects
correlations on the slow bond by dividing the system into two
homogeneous ones coupled at the defect site. Therefore this approach
is limited to single defect sites  ($l=1$). Comparison with MC results
shows a good agreement, at least for the high- and low-density phases.
The mean field treatment also leads to deviations of density profiles
near the defect site from the MC results.

In \cite{chou}, Chou and Lakatos gave an analytical approach treating 
a finite number of bottlenecks of arbitrary lengths. 
Their \emph{finite segment mean field theory (FSMFT)}  
divides the system into segments which contain one or a few defect sites.
By determining the leading eigenvector of the corresponding transition matrix
and matching the currents of the different subsystems, predictions for
the current through the system are obtained which at the same time
take into account correlations near the bottlenecks.
Although the method allows to treat several defect sites of arbitrary 
hopping rates within the segment, it is restricted to segment lengths
of less than 20 sites due to the numerical complexity. 

The results for one and two defect sites ($l=1$) in the bulk were 
recently extended by Dong et al.\ \cite{dong}. 
Additionally they have considered the case where a single defect is 
located near the boundary of the system and
found an \emph{edge effect} induced by the interaction of this defect
with the boundary. This implies a dependence of the current on the
position of the defect. Finally, the effects of Langmuir kinetics,
i.e.\ particle-creation and -annihilation in the bulk have been shown
\cite{frey2} to lead to a rich phase diagram including novel phases.

The present investigation generalizes the previous works in several
aspects. Especially we will study systematically the dependence of the
current on the length $l$ of the bottleneck and its position.
For this purpose we develop an analytical approach to calculate the current 
and critical boundary rates, called the \emph{interacting subsystem 
approximation (ISA)}. Mathematically it requires finding a specific
root, constrained by physical requirements, of a polynomial which has a 
degree of order $l$. 
For longer bottlenecks this approach is more efficient than the approach 
of \cite{chou} which demands to search an eigenvector of a 
$2^l\times2^l$-matrix.  
This allows us to generalize the single defect results of Dong et 
al.~\cite{dong} 
to longer bottlenecks. The increase of the maximum current by moving 
defects near the boundary is reproduced by ISA and extended to the case
$l>1$ (\emph{positive edge effect}).
However, for lower entry rates ISA predicts a decrease of the current
(\emph{negative edge effect}) which is confirmed by MC simulations.
We also observe in MC simulations an intermediate regime which shows a
non-monotonic dependence of the current on the position of the
bottleneck.  Furthermore we conclude that for bottlenecks near the
boundaries there is no phase transition to a maximum current phase,
since the current is not constant in any regime, but it approaches the
maximum current asymptotically for higher entry and exit rates,
and there is no macroscopic phase separation.

We also consider the case of two bottlenecks of arbitrary lengths and
separation generalizing the corresponding results of \cite{dong} 
for two single defects.  Therefore we also investigate the case that
one of the bottlenecks is near the boundaries. The results will
motivate a concept of \emph{effective boundary rates} that encompasses
the effect of boundary-near bottlenecks. This provides the basis for
an extensive study \cite{inprep} of finite defect densities, i.e.\ a
macroscopic number of slow bonds. Here previous investigations for
periodic \cite{barma,barma2,krug,harris1} and open systems
\cite{harris1,kolw,derrida2,lakatos,juhasz,foulaad} have revealed
surprising results.  For the open system, for example, it has been
found \cite{derrida2} that the position of the phase transitions is
sensitively sample-dependent even for large systems.

%%%%%%%%%%%%%%%%%%%%%%%%%%%%%%%%%%%%%%%%%%%%%%%%%%%%%%%%%%%%%%%%%%%%%%%%%%%%%%
%%%%%%%%%%%%%%%%%%%%%%%%%%%%%%%%%%%%%%%%%%%%%%%%%%%%%%%%%%%%%%%%%%%%%%%%%%%%%%
\section{Definition of the model}
\label{sec-analyt}

We consider a TASEP consisting of $L$ sites which can either be empty
or occupied by one particle. Throughout this paper we are mainly
interested in the steady-state in the limit $L\to\infty$.
With each bond between neighbouring sites
$j$ and $j+1$ we associate a hopping rate $p_{j,j+1}$ which
corresponds to the rate at which a particle will move from $j$ to its
right neighbour $j+1$ if this is empty\footnote{Sometimes it is more
  convenient to associate hopping rates $p_j$ with sites $j$. These
  site rates are related to the bond rates by $p_j=p_{j,j+1}$.  We
  will use both terminologies here.}.  At the boundary sites $j=1$ and
$j=L$ particles can be inserted and removed, respectively. If site $1$
is empty a particle will be inserted there with rate $\alpha$. On the
other hand, if site $L$ is occupied this particle will be removed with
rate $\beta$.  Here we will use a random-sequential update
corresponding to continuous-time dynamics.
%%%%%%%%%%%%%%%%%%%%%%%%%%%%%
\begin{figure*}[ht]
\begin{center}
  \vspace{0.5cm}
  \includegraphics[height=5cm]{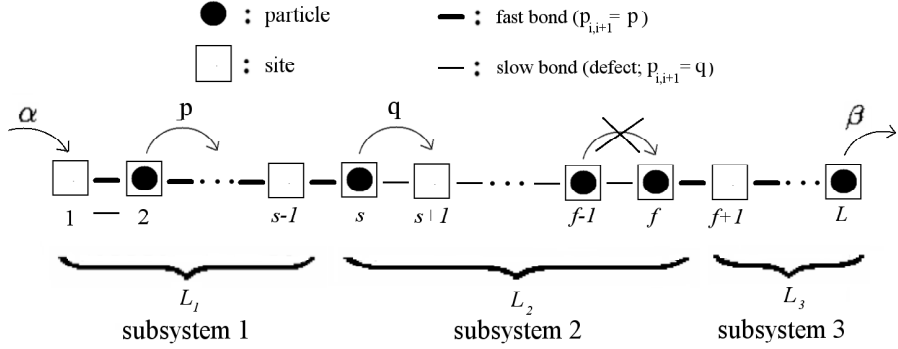}
\caption{\label{fig-asepdef}
  Illustration of the TASEP with a bottleneck of length $l$, i.e.\ $l$
  consecutive defect bonds (or sites resp.). The defect bonds/sites
  are characterized by a hopping rate $q<1$, whereas at the other
  bonds/sites the hopping rate is $p=1$. The system is virtually
  divided into three subsystems with virtual entry/exit rates
  corresponding to the mean occupation/vacancy on the adjacent sites.}
\end{center}
\end{figure*}
%%%%%%%%%%%%%%%%%%%%%%%%%%%%%%%%

In this paper we study the effect of a single bottleneck of length
$l$, i.e.\ a section of $l$ consecutive slow bonds (or sites)
with hopping rate $p_{j,j+1}=q<1$ (Fig.~\ref{fig-asepdef}).  All other
bonds are fast bonds with hopping rate $p_{j,j+1}=1$. 

In order to determine relevant quantities like the current and
critical boundary rates, we approach the problem by virtually dividing
the system into three interacting subsystems (Fig.~\ref{fig-asepdef}):
A system of $L_1$ sites with hopping rate $p_j=1$ at the left end of
the chain, a section of length $L_2=l+1$ consisting of the bottleneck
with $l$ slow sites (hopping rate $q$) extending from site $s=L_1+1$
to site $L_1+l=L_1+L_2-1$ plus an extra fast site $f=L_1+L_2$ at the
right end, and finally another pure system of length $L_3$ with
hopping rates $p_j=1$.  Note that subsystem 2, which contains the
bottleneck, starts and ends with a fast bond. This symmetry of the
subsystem will be more convenient for our investigations.

The main idea of our approach is to use the exact solution for the
stationary state of the pure TASEP with random-sequential dynamics 
\cite{derridaEHP,schuetz}.
This solution applies to all three pure subsystems defined above.  The
interactions between these subsystems are described by suitably chosen
boundary rates.
We neglect correlations at the sites connecting the subsystems.
Introducing the occupation number $\tau_j=0,1$ of site $j$, this
explicitly means that we assume 
\begin{equation}
\langle\tau_{j}(1-\tau_{j+1})\rangle \approx \rho_j(1-\rho_{j+1})
\qquad {\rm for\ }j=s-1,f
\end{equation}
where $\rho_j:=\langle \tau_j\rangle$ is the average local density 
at site $j$.
This approach is similar in spirit to that of Kolomeisky
\cite{kolomeiski2} for the case of a single defect ($l=1$). However, there
correlations on the slow bond are neglected whereas
here we will treat the bottleneck also as a pure system
(of reduced hopping rate) so that the most relevant correlations
induced by it are taken into account.  The three subsystems are
coupled through virtual boundary rates $\alpha_j$, $\beta_j$. Note
that all these rates are associated with fast bonds.
In the following we will call this approach the 
{\em interacting subsystem approximation (ISA)}  to distinguish it from 
the usual mean-field approach which neglects {\em all} correlations.

The rate equations at the sites connecting the subsystems are then
given by
\begin{eqnarray}
\label{bound_rates1}
  \frac{d}{dt}\langle \tau_{s-1}\rangle=\langle \tau_{s-2}
  (1-\tau_{s-1})\rangle-\beta_1\langle \tau_{s-1}\rangle \,, \\ 
  \frac{d}{dt}\langle \tau_{s}\rangle=\alpha_2(1-\langle\tau_{s}
  \rangle)-q\langle \tau_{s}(1-\tau_{s+1})\rangle\,,
\end{eqnarray}
and
\begin{eqnarray}
  \frac{d}{dt}\langle \tau_{f}\rangle&=&q\langle \tau_{f-1}
  (1-\tau_{f})\rangle-\beta_2\langle \tau_{f}\rangle \,,\\ 
  \frac{d}{dt}\langle \tau_{f+1}\rangle&=&\alpha_3(1-\langle\tau_{f+1}
  \rangle)-\langle \tau_{f+1}(1-\tau_{f+2})\rangle\,,\label{bound_rates2}
\end{eqnarray}
with the virtual boundary rates
\begin{eqnarray}
\label{def_bound_rates}
  \alpha_2 = \rho_{s-1}\,,\qquad \alpha_3 = \rho_{f}\,,\qquad \beta_1 =
  1-\rho_{s} \,,\qquad \beta_2 = 1-\rho_{f+1}\,,
\end{eqnarray}
in terms of the average local density $\rho_j$.
For completeness, we also define $\alpha_1:=\alpha$
and $\beta_3:=\beta$.  
Note that the mean-field factorization of
expectation values only applies to the sites at the boundaries of the
subsystems. All correlations within the subsystems will be taken into
account exactly! 

From the exact solution of the TASEP with $N$ sites for hopping rate
$p=1$, we know the ``partition function'' \cite{derridaEHP}
\begin{eqnarray}
\label{Z(N)}
\mathcal{Z}(\alpha,\beta,N) = \sum_{\lbrace\tau_i=0,1\rbrace}
\mathcal{P}(\lbrace \tau_i\rbrace) = \sum_{j=1}^N
\frac{j(2N-1-j)!}{N!(N-j)!} \frac{(1/\beta)^{j+1}
  -(1/\alpha)^{j+1}}{(1/\beta)-(1/\alpha)} ,\quad\,
\end{eqnarray}
which is exact for all system sizes $N\geq 1$. 
Here $\mathcal{P}(\lbrace \tau_i \rbrace)$ is the probability of finding 
the stationary system in a configuration $\lbrace \tau_i\rbrace$.
The singularity at $\alpha = \beta$ can be removed by taking the limit
$\alpha\to \beta$
which yields the analytic continuation
\begin{eqnarray}
  \mathcal{Z}(\alpha,N) :=
  \lim_{\beta\to\alpha}\mathcal{Z}(\alpha,\beta,N) = \sum_{j=1}^N
  \frac{j(2N-1-j)!}{N!(N-j)!}(j+1)\left( \frac{1}{\alpha}\right)^j \,
  .
\label{Z(N)2}
\end{eqnarray}

The corresponding results for the TASEP with hopping rate $p\neq 1$
can be obtained by rescaling of the boundary rates
$\alpha\to\alpha/p,\, \beta\to\beta/p$.  Thus the partition function
in a pure TASEP with hopping rate $p$ is
\begin{equation}
  \mathcal{Z}_p(\alpha,\beta,N)=\mathcal{Z}\left(\frac{\alpha}{p},
    \frac{\beta}{p},N\right) \,.
\label{Zp(N)}
\end{equation}
This result will be used in the following for different values of $N$,
depending on the length of the subsystems.

The exact current $J_0$ of a pure system of length $N$, which is the
quantity most relevant in the following, can be expressed through the
partition function as~\cite{derridaEHP}
\begin{equation}
\label{J(N)}
J_0(\alpha,\beta,N)=
\frac{\mathcal{Z}(\alpha,\beta,N-1)}{\mathcal{Z}(\alpha,\beta,N)}\, .
\end{equation}

Conservation of current then yields the central equations of the ISA 
\begin{equation}
\label{current_cons}
J_0(\alpha_1,\beta_1, L_1)=
J_0\left(\frac{\alpha_2}{q},\frac{\beta_2}{q},L_2\right)
=J_0(\alpha_3,\beta_3,L_3)
\end{equation}
which express the fact that the current in the stationary state is
the same in all three interacting subsystems. 
Note that we have used (\ref{Zp(N)}) and (\ref{J(N)}) to express the
current in the bottleneck (subsystem 2) through the result $J_0$ for a
pure system.

Inserting (\ref{Z(N)}) and (\ref{J(N)}) in (\ref{current_cons}) we
obtain a set of two algebraic equations with six variables.  In
principle one can solve these equations, if four of the variables are
known or more equations are given that determine variables.

%%%%%%%%%%%%%%%%%%%%%%%%%%%%%%%%%%%%%%%%%%%%%%%%%%%%%%%%%%%%%%%%%%%%%%%%%%%%%%
%%%%%%%%%%%%%%%%%%%%%%%%%%%%%
\section{One bottleneck far from the boundaries}
\label{one_bottleneck}

First we study the case where the bottleneck is
\emph{far from the boundaries}. So we assume that $L_1$ and $L_3$ are large.
From the latter condition it is expected that the topology of the
phase diagram is the same as that of a system with a single defect
($l=1$). We can therefore follow \cite{kolomeiski2} and classify the
phases by the phases of the pure subsystems 1 and 3 which are
${\mathcal{O}}(L)$.

Though at first glance one could expect nine possible phases
corresponding to all possible combinations of three phases l,h,m (see
Sec.~\ref{introduction}) that can be realized in the pure subsystems 1
and 3, it was argued in \cite{kolomeiski2} that only the combinations
l-l, h-h and h-l can exist. The l-l-phase corresponds to a global 
\emph{low density phase (L)}, while h-h corresponds to a 
\emph{high density phase (H)}.
In both cases, only around the bottleneck there are local deviations
from the density profile of the homogeneous system. In the h-l-phase,
we have phase separation which cannot be observed in the pure system.
The current in the h-phase is independent of the entry rate, while in
the l-phase it is independent of the exit rate. Thus in a phase separated 
h-l-phase, the current is independent of both boundary conditions 
and takes a maximum value (see below). 
Although it is sometimes called a \emph{maximum current phase (M)} like in 
the pure system, its properties differ from
the maximum current phase in the pure system not only by the
occurrence of phase separation, but also by the absence of algebraic
boundary layers. Therefore we prefer the terminology \emph{phase
  separated regime} (PS).  Furthermore, the transition to this phase
corresponds to a transition of subsystem 1 from l to h and vice versa
respectively, which is accompanied by a discontinuity of the mean
density $\langle \rho \rangle=\sum_j \rho_j$.
According to this, it can be classified as a first order transition in
contrast to the pure system where the transition to the maximum
current phase is of second order.

Though the system with a bottleneck is not exactly invariant under the
\emph{particle-hole symmetry operation} defined by $\tau_j\leftrightarrow
1-\tau_j,\, \alpha\leftrightarrow \beta, \,j\leftrightarrow L-j$, this
transformation only changes the position of the bottleneck, but still
leaves it far from the boundary. Thus the particle-hole transformation
leaves the phases of the subsystems unchanged, since it only changes
their sizes, but they stay ${\mathcal{O}}(L)$. Therefore, we can
conclude that the phase diagram must be symmetric with respect to the
line $\alpha=\beta$. This symmetry constraint yields that the
transition line between high and low density phase must be at
$\alpha=\beta$.

We now want to determine the critical entry rate $\alpha^*$ at the
transition from the L-phase to the PS-phase for fixed $\beta$ in terms
of the analytical ISA approach introduced in the last section.  It
corresponds to the transition of (the pure) subsystem~1 from low to
high density phase which occurs at $\alpha^*=\alpha_1=\beta_1$.
Furthermore we know that at this point $J=\alpha_1(1-\alpha_1)$.
Subsystem 3 still remains in the low density phase. Therefore we
also have $J=\alpha_3(1-\alpha_3)$. Since the current must be the same
as in subystem 1 and $\alpha_3$ must be smaller than 1/2 we conclude
$\alpha_3=\alpha^*$. From the definition of the virtual boundary rates
(\ref{bound_rates1}), (\ref{def_bound_rates}) we obtain by a simple
transformation
\begin{eqnarray}
\beta_2 &=& J/\rho_{f}=J/\alpha^*=1-\alpha^*\,, \\
\alpha_2 &=& J/(1-\rho_{s})=J/\beta_1=J/\alpha^*=1-\alpha^* \, ,
\end{eqnarray}
where the first equality in each equation can be found e.g.\ in 
\cite{derridaEHP}.
From the exact solution of the pure TASEP (\ref{J(N)}), (\ref{Z(N)})
and the conservation of current (\ref{current_cons}), it follows
\begin{eqnarray}
\label{a*-equ}
&\,& J\left(\frac{1-\alpha^*}{q},\frac{1-\alpha^*}{q},l+1\right)
     =\alpha^*(1-\alpha^*) \nonumber \\ 
\Longleftrightarrow \hspace{5mm} && \alpha^*(1-\alpha^*)\mathcal{Z}
((1-\alpha^*)/q,l+1)= \mathcal{Z}((1-\alpha^*)/q,l) \, ,
\end{eqnarray}
where $\mathcal{Z}(1-\alpha,l+1)$ is the limit defined in
(\ref{Z(N)2}) and $\mathcal{Z}(1-\alpha,l+1)\neq 0$ for
$0\leq\alpha\leq 1$.

Eq.~(\ref{a*-equ}) is essentially a polynomial in $1/\alpha^*$ (or
$\alpha^*$, respectively) and can be solved numerically (or
analytically for small values of $l$).  
The requirement that a physical relevant solution for $\alpha^*$ has
to be in the interval $[0,\frac{1}{2}]$ gives a unique solution
$\alpha^*(q,l)$ for the transition point to the PS-phase. 
In general, the solution depends on the length $l$ of the bottleneck
and the slow hopping rate $q$. 
For $l=1$, for example, equation (\ref{a*-equ}) can be transformed into
\begin{equation}
2{\alpha^*}^2-(2+3q)\alpha^*+2q=0 \, .
\end{equation}
The relevant solution is then 
\begin{equation}
\alpha^*=\frac{2+3q}{4} - \sqrt{\frac{(2+3q)^2}{16} - q}\,.
\label{a*(l=1)}
\end{equation}
Explicitly, for the value $q=0.6$ used in most simulations,
one obtains $a^*=\frac{2}{5}$ by evaluating (\ref{a*(l=1)}).

The current at the transition point corresponds to the maximum current
$J^*$ in the PS-phase which can be interpreted as the \emph{transport
  capacity} of the system. From (\ref{a*-equ}) we can see that
\begin{equation}
\label{Jmax-equ}
J^{*}(q,l)=\alpha^*(q,l)(1-\alpha^*(q,l)) \, ,
\end{equation}
which yields $J^*(0.6,1)=\frac{6}{25}$ for $q=0.6$
after taking into account (\ref{a*(l=1)}).

\begin{figure}[ht]
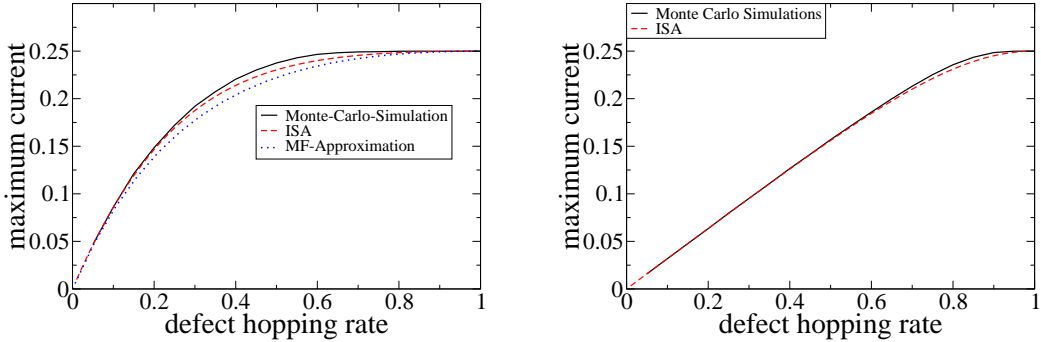

\begin{center}
  \vspace{0.5cm}
  \includegraphics[width=0.46\columnwidth]{Jmax_q_Ld=1.eps} \qquad
  \includegraphics[width=0.46\columnwidth]{Jmax_q_Ld=6.eps}
\end{center}
\caption{\label{Jmax_q}Maximum current in a system with a bottleneck far from the 
  boundaries in dependence on the slow hopping rate $q$. It is
  determined by Monte Carlo simulations for $\alpha=\beta=0.5$ and
  by solving the ISA equation (\ref{a*-equ}) for a bottleneck of length
  $l=1$ (left) and $l=6$ (right). For comparison the mean field
  approximation $J^*_{{\rm MF}}=\frac{q}{(1+q)^2}$ for $l=1$ is
  included.}

\end{figure}

In Fig.~\ref{Jmax_q} we have plotted the dependence of $J^*(q,l)$ on
the slow hopping rate $q$ (for fixed $l$) obtained by (\ref{Jmax-equ})
and compare these analytical ISA-results with results from Monte Carlo
(MC) simulations for different bottleneck lengths. For a single defect
site ($l=1$) we have compared the analytical results obtained by
(\ref{a*(l=1)}) and (\ref{Jmax-equ}) with the results obtained by
pure mean field approximation $J^*_{MF}=\frac{q}{(1+q)^2}$
\cite{kolomeiski2}.  Obviously also for $l=1$ the results obtained by
(\ref{Jmax-equ}) are more accurate than mean field results, while for
longer bottlenecks, to our knowledge, no proper mean field
approximations are known. Note that our approximation takes into
account correlations on the slow bonds, only correlations on sites
adjacent to the bottleneck are neglected.
%%%%%%%%%%%%%%%%%%%%%%%%%%%%%%%%
%%%%%%%%%%%%%%%%%%%%%%%%%%%%%%%%%%%

\begin{figure}[ht]
\begin{center}
  \vspace{0.65cm}
  \includegraphics[width=0.55\columnwidth]{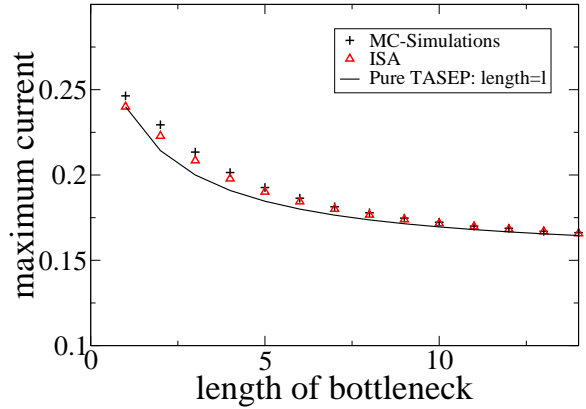}
\end{center}
\caption{\label{Jmax(l)}Maximum current in dependence of the bottleneck size for $q=0.6$. 
  MC-simulations are compared with ISA results. The solid line
  is the current in a homogeneous TASEP with $L=l+1$,\,
  $\alpha=\beta=1$ and hopping rate $q=0.6$.}
\end{figure}

Fig.~\ref{Jmax(l)} shows the maximum current in dependence of the
bottleneck length $l$. Here we compare MC-simulations with the results 
obtained by (\ref{Jmax-equ}). The ISA results
systematically underestimate the real current. The deviation is
largest for small $l$, but it does not exceed 3\%. For larger
bottlenecks, the agreement improves. We also note that we have
chosen $q=0.6$ since for this value the observed deviations have been
found to be largest. In addition we plotted the exact
current of the pure system with system size $L=l+1$ and
$\alpha=\beta=1$. It seems that the asymptotics for large $l$ are the
same as for the bottleneck.
%%%%%%%%%%%%%%%%%%%%%%%%%%%%%%%%%%%%
%%%%%%%%%%%%%%%%%%%%%%%%%%%%%%%%%%%

\begin{figure}[ht]
\begin{center}
  \vspace{0.2cm}
  \includegraphics[width=0.47\columnwidth]{mass+J_a_q=0.6_Ld=5.eps}
  \quad \includegraphics[width=0.44\columnwidth]{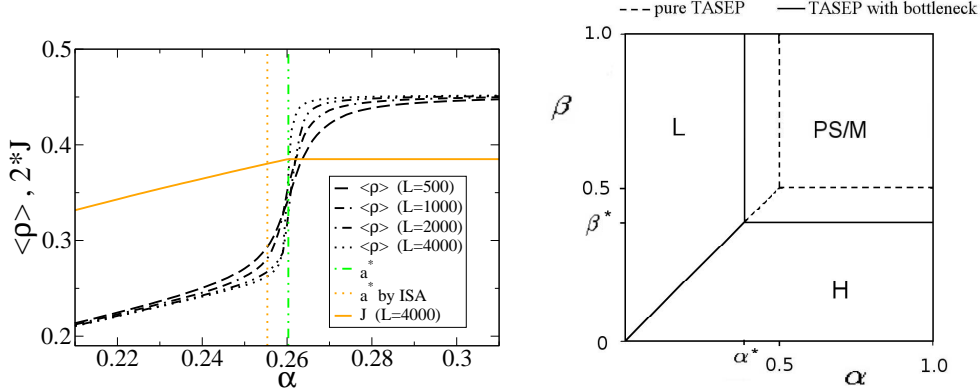}
\end{center}
\caption{\label{fig-phase}Left: Mean density $\langle \rho \rangle$ 
  and current $J$ as function of $\alpha$. The slope of $\langle \rho
  \rangle(\alpha)$ becomes steeper for larger systems indicating a
  discontinuity in the thermodynamic limit. The vertical lines show
  the steepest point of $\langle \rho \rangle(\alpha)$ identified as
  the phase transition $\alpha^*$ compared with the ISA results from
  (\ref{a*-equ}).\newline Right: Schematic phase diagram of the TASEP
  with one bottleneck far from the boundaries. It looks similiar to
  the phase diagram of the pure TASEP while the transition lines to
  the maximum current (phase separated) phase is shifted to be at
  $\alpha^*$ and $\beta^*$. The mean density is discontinuous at these
  points. }
\end{figure}

In Fig.~\ref{fig-phase} the dependence of the mean density $\langle
\rho \rangle=\frac{1}{L}\sum_{i=1}^{L} \rho_i$ and the current $J$ on
the entry rate $\alpha$ is plotted for fixed $\beta$.  
In order to compute these quantities, we used the
method introduced in \cite{derrida2} that allows to calculate the
current and densities for an arbitrary set of values of $\alpha$ in
one simulation. One observes a steep increase of the mean density for
the same value of $\alpha$ where the maximum current plateau begins.
This seems to coincide with the value of $\alpha^*$ obtained
by ISA quite well. The slope increases with system size
indicating a discontinuity in the thermodynamic limit corresponding to
a first order phase transition, in contrast to the pure TASEP. The
plots also show that at the transition point and in the hole PS-phase,
the current is maximal.

As we have argued above, the phase diagram must be symmetric with
respect to the diagonal $\alpha=\beta$ that yields
$\beta^*=\alpha^*$.
With this information we can sketch the phase diagram of the TASEP
with one bottleneck far from the boundaries displayed in
Fig.~\ref{fig-phase}. Its topology is the same as the one for a single
defect far from the boundaries \cite{kolomeiski2}, while longer
bottlenecks have a larger PS-phase than single defects. It also looks
similiar to the phase diagram of the pure TASEP but here all phase
transitions are of first order in contrast to the TASEP and the
characteristics of the M-phase and PS-phase are different.
%%%%%%%%%%%%%%%%%%%%%%%%%%%%%%%%

%%%%%%%%%%%%%%%%%%%%%%%%%%%%%%%%

Another procedure to compute the maximum current for a finite
bottleneck is the \emph{finite segment mean field theory (FSMFT)}
introduced in \cite{chou}. In this approach, a segment of $l'$ sites
including the bottleneck of length $l < l'$ is considered.
Currents can then be obtained from the eigenvector of the
zero-eigenvalue of the  $2^{l'}\times 2^{l'}$-transition matrix
of this segment \cite{chou}.
The advantage of this method is that the accuracy can be systematically
increased by expanding the size $l'$ of the segment. It can also treat
arbitrary combinations of hopping rates inside the segment.
However, due to the exponentially decreasing size of the transition
matrix one is currently restricted to segment lengths $l< 20$.
In contrast, the ISA-method, though not asymptotically exact, relies
on finding \emph{one specific} root of a polynomial equation of a maximal
degree $l+2$.  
This makes it possible to compute the maximum current for systems with 
large bottlenecks of several hundred sites rather easily. 
This advantage becomes relevant for disordered systems with finite
defect site density where bottlenecks of arbitrary length can occur.
For example the computation of the maximum current of a system with 500
consecutive defect sites can be made in less than one second on a
standard\footnote{AMD Athlon 3000MHz} PC to obtain $J^*_{ISA}(l=500)=0.15045$
for $q=0.6$. Although this agrees nicely with the value from MC simulations, 
$J^*_{MC}=0.15049$, it is clearly different from the asymptotic value 
$J^*_{\infty}=q/4=0.15$ for $l\to\infty$ \cite{krug,barma,chou}. 

%%%%%%%%%%%%%%%%%%%%%%%%%%%%%%%%%%%%%%%%%%%%%%%%%%%%%%%%%%%%%%%%%%%%%%%%%%%%%%
\section{Edge effects: One bottleneck near a boundary}
\label{one_bottleneck_at_boundary}

Next we consider a system with a bottleneck near the left boundary,
i.e.\ now both $L_1$ and $L_2=l+1$ are of order $\mathcal{O}(1)$
for $L\to\infty$.
The case of a bottleneck at distance $d$ from the right boundary,
i.e.\ the last slow site is at site $L-d-1$, has not to be considered
separately since we can deduce the results using the particle
hole-symmetry
\begin{eqnarray}
  && \alpha\leftrightarrow\beta, \,\,\, \rho_j\leftrightarrow
  1-\rho_j\,, \,\,\, j\leftrightarrow L-j, \,\,\, \mbox{subsystem
    1}\leftrightarrow \mbox{subsystem 3} \,.
\label{p-h-symmetry}
\end{eqnarray}

Since we have only one macroscopic subsystem, namely subsystem 3, the
classification of phases is slightly different than in
Sec.~\ref{one_bottleneck}. Now the phase of the system with bottleneck
is basically identically to that of subsystem 3.  Phase separation can
no longer occur since the size of subsystem 1 is microscopic.  The
entry rate $\alpha_3$ of subsystem 3, defined in ISA (see
(\ref{bound_rates1})-(\ref{bound_rates2})), can thus be treated as an
\emph{effective entry rate} $\alpha_{\mathrm{eff}}:=\alpha_3\neq
\alpha$ for the {\em bulk} of the system which is a homogeneous TASEP.
The phase of the full system corresponds to the phase of subsystem 3
which we now denote by L', H' or M.

For $L_1\geq 2$, we can divide the system into three subsystems in the same 
manner as in Fig.~1. First we consider a system in the
low-density phase L'. 
Therefore the current is given by 
\begin{equation}
\label{J(aeff)}
J=\alpha_{\mathrm{eff}}(1-\alpha_{\mathrm{eff}}) \, .
\end{equation}
In the steady state the entry and exit rates of a pure TASEP with $N$
sites are $\alpha=\frac{J}{1-\rho_{1}}$ and $\beta=\frac{J}{\rho_{N}}$ 
\cite{derridaEHP}.  The currents in subsystems 1 and 2 have to be identical 
and thus, as in (\ref{current_cons}),
\begin{eqnarray}
\label{current_cons(near_bound)_1}
J_0(\alpha,1-\rho_{s},L_1) &=& J_0\left(\frac{J}{q(1-\rho_{s})},
\frac{J}{q\rho_{f}},L_2\right)\,, 
\end{eqnarray}
where $J_0$ is the exact current in a pure system given by
(\ref{J(N)}) and $\rho_s$ and $\rho_f$ are defined in the same
manner as in Sec.~\ref{one_bottleneck}.  Since subsystem 3 is
assumed to be in the low density phase, the density profile is flat
at its left end, i.e.\ the density is independent of the position 
if we apply ISA.
Therefore we have $\alpha_{\mathrm{eff}}=\rho_f=\rho_{f+1}$ and can take
$J=\rho_f(1-\rho_f)$ in eq.~(\ref{current_cons(near_bound)_1}). 
Furthermore the currents in subsystem 1 and subsystem 3 have to be the 
same which leads to 
\begin{eqnarray}
\label{current_cons(near_bound)_2}
J_0(\alpha,1-\rho_{s},L_1) = \rho_{f}(1-\rho_{f}) \, .
\end{eqnarray}
Now we have two ISA-equations, (\ref{current_cons(near_bound)_1}) and
(\ref{current_cons(near_bound)_2}), as well as the two variables
$\rho_s$ and $\rho_f=\alpha_{\mathrm{eff}}$. For given $\alpha$, $L_1$ and
bottleneck length $l$ these equations can be solved to obtain the
effective entry rate $\alpha_{\mathrm{eff}}$, while the resulting current 
can be obtained from (\ref{J(aeff)}).

For $L_1=1$ that procedure does not work in the same way because the 
current in a system of size 1 is not defined as one can see in (\ref{J(N)}). 
But even in this case, we still have $\alpha=J/(1-\rho_1)$ which is 
valid by definition of the boundary rates. 
After inserting $J=\rho_f(1-\rho_f)$, this equation together 
with (\ref{current_cons(near_bound)_2}) again gives a solvable set of 
two equations with the two variables $\rho_s$ and $\rho_f$.

For $L_1=0$ subsystem 1 does not exist and we have only the
two subsystems 2 and 3.  Subsystem 2 comprises the sites $\lbrace
1,\ldots,L_2\rbrace$ and subsystem 3 $\lbrace L_2+1,\ldots,L\rbrace$ 
which includes the bulk of the system. This problem can however be solved,
by inserting 
\begin{eqnarray}
\alpha_2=\alpha\,, \qquad 
\beta_2=\frac{J}{1-\rho_f}\,, \qquad
\alpha_3=\rho_f\,.
\end{eqnarray}
Analogous to (\ref{current_cons}) we thus obtain the equation
\begin{equation}
\label{aeff-equ_d=0}
J_0\left(\frac{\alpha}{q},\frac{J}{q(1-\rho_f)},L_2\right) 
= \rho_f(1-\rho_f) \,,
\end{equation}
with $J= \rho_f(1-\rho_f)$. This equation can be solved for the
variable $\rho_f=\alpha_{\rm eff}$ thus we obtain the effective entry rate
and by (\ref{J(aeff)}) the corresponding current.

Because of the particle-hole symmetry (\ref{p-h-symmetry}) we can transfer 
these results for $\beta_{{\rm eff}}$ in the high-density phase. 
The effective exit rate for the bulk $\beta_{{\rm eff}}$ can then
be determined using (\ref{J(aeff)})-(\ref{aeff-equ_d=0}).
Note that in the low-density phase, $\beta$ has no influence on the
bulk of the system, only in a small region near the boundaries. The same is
valid in a high-density phase for $\alpha$. 

%%%%%%%%%%%%%%%%%%%%%%%%%%%%%%%%%%%%%%%%%%%%%%%%%%%%%%%

\begin{figure}[ht]
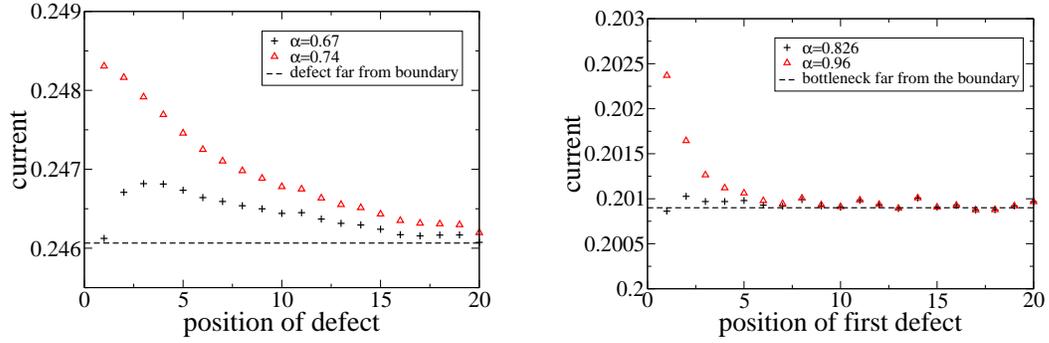

\begin{center}
\vspace{0.5cm}
\includegraphics[width=0.46\columnwidth]{J_m+p_edgeeffect_l=1.eps}
\qquad
\includegraphics[width=0.46\columnwidth]{J_m+p_edgeeffect_l=4.eps}
\end{center}
\vspace{0.6cm}
\caption{\label{J_bound_Ld}
  Current in a system with a single defect site (left) and a
  bottleneck of length $l=4$ (right) near the left boundary in
  dependence on its position, for high entry rates. The exit rate is
  $\beta=1.0$.  The system size is $L=500$ while the number of
  iterations is 40000000, $q=0.6$. The entry rates have been chosen in
  order to optimally display the mixed and the positive edge effect.
  In the system with the longer bottleneck (right), the mixed edge
  effect is almost not visible even for optimized parameters. Note
  that the magnitude is much smaller than in Fig.~\ref{J_bound_Ld_2}.
  }
\end{figure}

%%%%%%%%%%%%%%%%%%%%%%%%%%%%%%%%%%%%%%%%%%%%%%%%%%
\begin{figure}[ht]
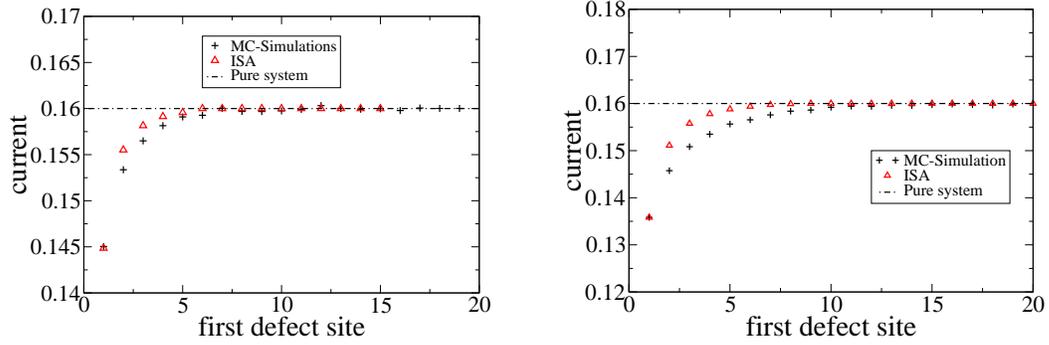

\begin{center}
\vspace{0.5cm}
\includegraphics[width=0.46\columnwidth]{J_a=0.2_Ld=1.eps}
\qquad
\includegraphics[width=0.46\columnwidth]{J_a=0.2_Ld=4.eps}
\end{center}
\vspace{0.6cm}
\caption{\label{J_bound_Ld_2}
  Current in a system with a bottleneck near the boundaries in
  dependence of the position of the first defect site, for low entry
  rates. The exit rate is $\beta=\alpha$, system size $L=500$ and
  number of MC steps $=2000000$ for bottleneck length $l=1$, $q=0.6$,
  and $\alpha=0.2$ (left), bottleneck length $l=4$, $q=0.6$ and
  $\alpha=0.2$ (right).  The scale for the current has been chosen
  in order to emphasize the form of the deviations.}

\end{figure}
%%%%%%%%%%%%%%%%%%%%%%%%%%%%%%%%%%%%%%%%%%%%%%%%%%%%%%%

%%%%%%%%%%%%%%%%%%%%%%%%%%%%%%%%%%%%%%%%%%%%%%%%%%%%%%%%%%%%%%%%%%%%%%%%%%%%%%

\subsection{Edge effect: discussion}

In Fig.~\ref{J_bound_Ld} and~\ref{J_bound_Ld_2} we have plotted the
current in dependence on the position of the first defect site for
different values of $\alpha$ while $\beta=1$. One observes that the
position of the bottleneck has a significant influence on the current.
This is called the \emph{edge effect}, first observed in \cite{dong}.
In that work only an increase of the current was observed if defects
approach the boundary (\emph{positive edge effect}) which can be seen
in Fig.~\ref{J_bound_Ld}.  In addition, ISA predicts a decrease of the
current for low entry rates.  This \emph{negative edge effect} is
confirmed by MC simulations (Fig.~\ref{J_bound_Ld_2}).  There is also
a region of entry rates, where the dependence of the current on the
bottleneck position is non-monotonic (\emph{mixed edge effect}, see
Fig.~\ref{J_bound_Ld})!  Simulations for different system sizes
indicate that this non-monotonic behaviour is not a finite-size
effect.  Nonetheless, as we can also see in the figures, the magnitude
of the positive and the mixed edge effect is much smaller than the one
of the negative edge effect. The ISA results obtained from the
equations in the last subsection confirm the existence of negative and
positive edge effect, while the mixed one is not. This indicates that
the mixed edge effect is caused by correlations at the edge of the
bottleneck. In Fig.~\ref{J_bound_Ld_2} we plotted the analytical
results for comparison. We did not display them in
Fig.~\ref{J_bound_Ld} since the deviations due to correlations are
larger than the positive/mixed edge effect itself. In the appendix it
is shown that the negative edge effect is predominant in regimes where
the current depends significantly on the boundary rates.

\begin{figure*}[ht]
\begin{center}
  \vspace{0.7cm}
\includegraphics[height=5cm]{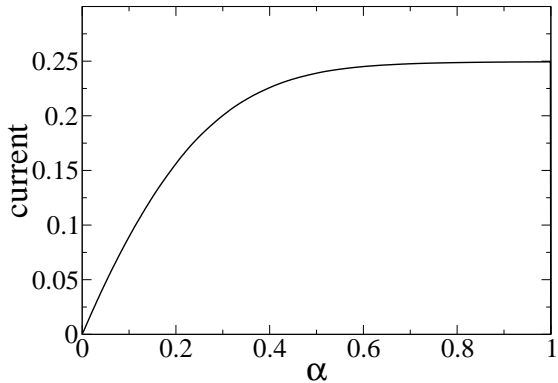}
\caption{\label{J(a)_bn_boundary}Current in dependence on the entry 
  rate $\alpha$ for $\beta=1$  and a defect at site $j=3$. Results are 
  obtained by MC simulations. There is no plateau. }
\end{center}

\end{figure*}

Moreover, we see that the current does not attain a plateau value.
Instead it seems to approach asymptotically the value $J=0.25$
(Fig.~\ref{J(a)_bn_boundary}). This is confirmed by our simulations,
where we calculated the current for the very high value $\alpha=100$
that yielded an effective entry rate of $\alpha_{\rm eff}\approx 0.5$.
Since we assume the effective entry rate to grow monotonic with the
real entry rate, we can conclude that we have always $\alpha_{\rm
  eff}<0.5$, thus the maximum current phase can never be attained. As
we have argued above, there is no phase separation, either, if the
distance of the bottleneck from the boundary is microscopic, so we can
state that for a bottleneck near the boundary only the high and low
density phase can occur. This is plausible since a similar argument
for the absence of the maximum current phase in subsystem~3 as in
Sec.~\ref{one_bottleneck} applies.

Though strictly speaking there is no M/PS-phase if the bottleneck is
near a boundary, the system still exhibits a kind of \emph{crossover}.
While there is neither a plateau region nor a sharp kink in the
dependence of the current on the entry rate (see
Fig.~\ref{J(a)_bn_boundary}), for higher entry rates this dependence
is rather weak. For a bottleneck far from the boundaries we actually
have sharp phase transitions, so we can say that by approaching the
bottleneck to the boundary, the phase transition to the M/PS-phase is
``softened'' into a crossover.

%%%%%%%%%%%%%%%%%%%%%%%%%%%%%%%%%%%%%%%%%%%%%%%%%%%%%%%%%%%%%%%%%%%%%%%%%%%%%%
%%%%%%%%%%%%%%%%%%%%%%%%%%%%%%%%%%%%%%%%%%%%%%%%%%%%%%%%%%%%%%%%%%%%%%%%%%%%%%

%%%%%%%%%%%%%%%%%%%%%%%%%%%%%%%%%%%%%%%%%%%%%%%%%%%%%%%%%%%%%%%%%%%%%%%%%%%%%%
\begin{figure*}[ht]
\begin{center}
  \vspace{0.7cm}

\includegraphics[height=5cm]{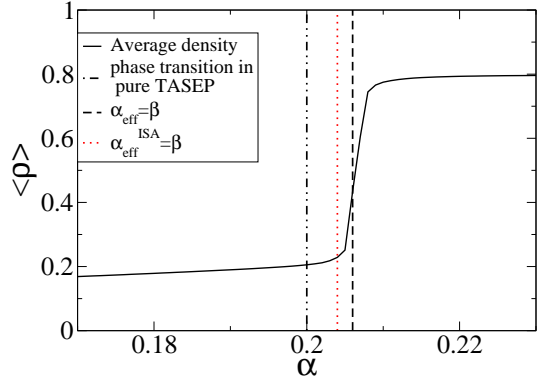}

\caption{\label{mass(a)_bn_boundary}Average density in dependence on 
  the entry rate for $\beta=0.2$ and a defect at site 3. The
  transition to the high density phase is shifted to larger $\alpha$
  compared to the pure system.}
\end{center}

\end{figure*}
%%%%%%%%%%%%%%%%%%%%%%%%%%%%%%%%%%%%%%%%%%%%%%%%%%%%%%%%%%%%%%%%%%%%%%%%%%%%%%

%%%%%%%%%%%%%%%%%%%%%%%%%%%%%%%%%%%%%%%%%%%%%%%%%%%%%%%%%%%%%%%%%%%%%%%%%%%%%%
\begin{figure*}[ht]
\begin{center}
  \vspace{0.7cm}
\includegraphics[height=5cm]{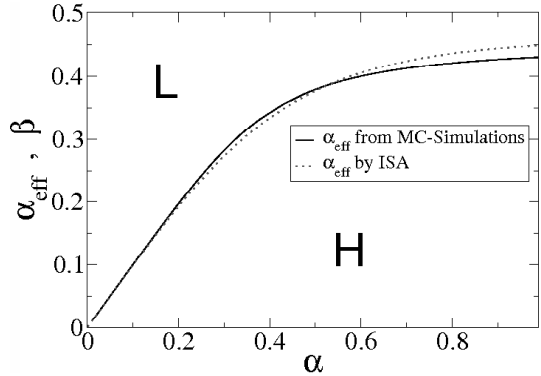}
\caption{\label{aeff(a)}Effective entry rate in dependence on $\alpha$. 
  Interpreting the y-axis as the beta range, this corresponds to the
  phase diagram with the graph being the transition line between
  H-phase (below) and L-phase (above).}
\end{center}
\end{figure*}
%%%%%%%%%%%%%%%%%%%%%%%%%%%%%%%%%%%%%%%%%%%%%%%%%%%%%%%%%%%%%%%%%%%%%%%%%%%%%%

The transition from low density phase to the high density phase occurs
for $\beta=\alpha_{\rm eff}$.  This is confirmed in
Fig.~\ref{mass(a)_bn_boundary} where we have plotted the average
density in dependence on $\alpha$. The jump in the average density
marks the transition point, which matches quite good the value of
$\alpha_{\rm eff}$.  One observes that the transition point is shifted
to higher values of $\alpha$ compared to the transition in the pure
system $\alpha=\beta$.  In this diagram, $\alpha_{\rm eff}$ was
obtained by the formula $\alpha_{\rm eff}=\frac{J}{1-\rho_{f+1}}$.
The value calculated by solving the ISA equations
(\ref{current_cons(near_bound)_1}) and (\ref{current_cons(near_bound)_2})
is a little less, but it still yields the correct sign
of the shift of the transition point.

Fig.~\ref{aeff(a)} shows the dependence of $\alpha_{\rm eff}$ on $\alpha$.
Since $\beta=\alpha_{\rm eff}$ is the transition line between high and low
density phase, the diagram simultaneously displays the phase diagram,
interpreting the y-axis as the $\beta$-range.
As we see the phase transition line calculated in ISA is in
good agreement with MC results.

%%%%%%%%%%%%%%%%%%%%%%%%%%%%%%%%%%%%%%%%%%%%%%%%%%%%%%%%%%%%%%%%%%%%%%%%%%%%%%
%%%%%%%%%%%%%%%%%%%%%%%%%%%%%%%%%%%%%%%%%%%%%%%%%%%%%%%%%%%%%%%%%%%%%%%%%%%%%%
\section{Two bottlenecks far from the boundaries}
\label{2bottlenecks_sect}

The next step towards disordered systems is the investigation of a
system with two bottlenecks.
This will tell us something about the importance
of "interactions" between the bottlenecks.
We distinguish two cases. First we will consider situations where
both bottlenecks are in the bulk of the system, i.e.\ far away from
the boundaries. Then we study edge effects in more detail by allowing
one bottleneck to be close to one of the boundaries.
Unfortunately, we cannot use the analytic ISA approach applied in the last 
sections since the equations would be underdetermined. 
Thus we have to rely on simulation results.

%%%%%%%%%%%%%%%%%%%%%%%%%%%%%%%%%%%%%%%%%%%%%%%%

\subsection{Two bottlenecks far from the boundaries}
\label{2bottlenecks_far}

We simulated systems with two bottlenecks of length $l_1$ and $l_2$ 
with $d$ fast sites in between.
We focussed on the maximum current phase and determined the current 
$J^*_{2}$. 
In Fig.~\ref{Jmax(L1,L2)}, $J^*_{2}$ is plotted as function
of the distance $d$ between the bottlenecks for different values 
of $l_1$ and $l_2$. 
One sees that, if the lengths of the two bottlenecks
differ with $l_1>l_2$, $J^*_{2}$ tends to converge to $J^*(l_1)$,
which is the value obtained in a system with only the longer
bottleneck. The convergence is faster for a larger difference  $l_1-l_2$
of the bottleneck lengths. In this case for a distance of
about 5-10 lattice sites, the maximum current is almost the same as
for a system with only the longer of the two bottlenecks. 
Of course for $d\to 0$ the bottlenecks merge, thus we have only one 
bottleneck and $J^*_{2}(l_1,l_2,0)=J^*(l_1+l_2)$. 
If both bottlenecks have equal size, $d\to\infty$ converges to the 
maximum current of a single bottleneck which
generalizes results of \cite{dong} to the case $l_1=l_2 > 1$.
%, investigating two single defect sites.

%%%%%%%%%%%%%%%%%%%%%%%%%%%%%%%%%%%%%%%%%%%%%%%%%%%%%%%%%%
\begin{figure}[t]
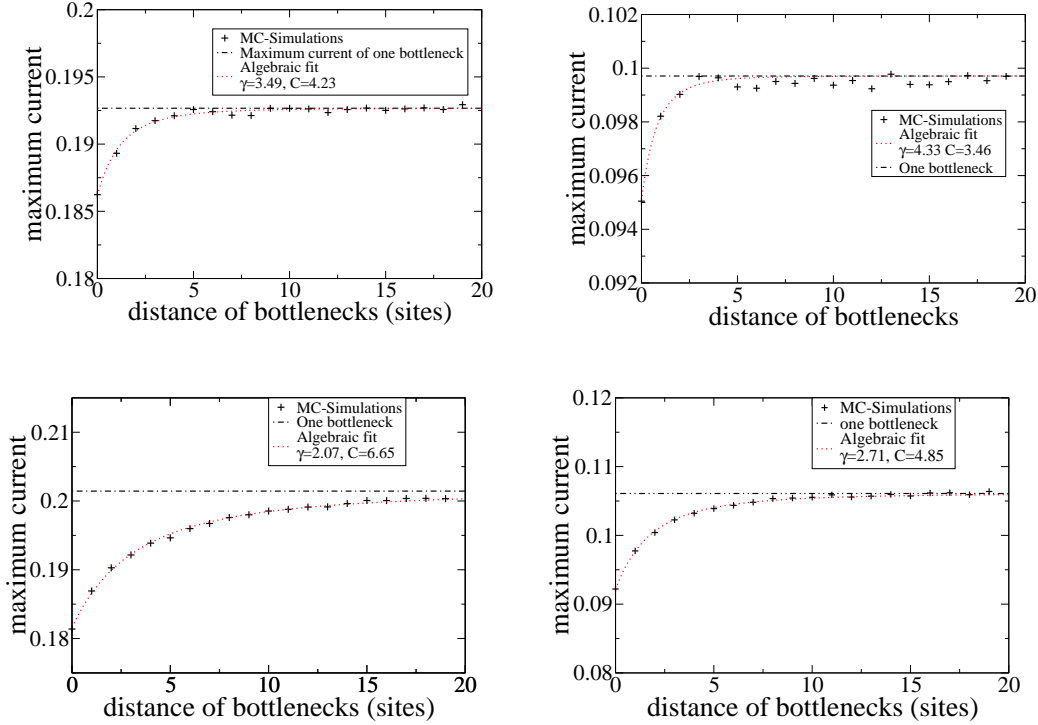

\begin{center}
\includegraphics[width=0.45\columnwidth]{Jmax_L1=5_L2=1_q=0.6.eps}
\qquad
\includegraphics[width=0.45\columnwidth]{Jmax_L1=5_L2=1_q=0.3.eps}
\end{center}
\vspace{0.8cm}
\includegraphics[width=0.45\columnwidth]{Jmax_L1=4_L2=3_q=0.6.eps}
\qquad
\includegraphics[width=0.45\columnwidth]{Jmax_L1=4_L2=3_q=0.3.eps}
\caption{\label{Jmax(L1,L2)}
  Maximum current in a system with two bottlenecks of length $l_1$
  (the first one) and $l_2$ (the second one) far from the boundaries
  in dependence on the distance of the two defects. The current is
  determined by MC simulations for $\alpha=\beta=0.5$. In each graph
  an algebraic fit of the form $f(d)=J^*(l_1)-(d+C)^{-\gamma}$ with
  fit parameters $C$ and $\gamma$ is included. $J^*(l_1)$ is obtained
  from the MC simulations in Sec.~\ref{one_bottleneck}. One observes,
  that the maximum current converges to the maximum current of a
  system with a single bottleneck, i.e. the longer one. The parameters
  are $l_1=5$, $l_2=1$, $q=0.6$ (top left), $l_1=5$, $l_2=1$, $q=0.3$
  (top right), $l_1=4$, $l_2=3$, $q=0.6$ (bottom left), and $l_1=4$,
  $l_2=3$, $q=0.3$ (bottom right).}

\end{figure}
%%%%%%%%%%%%%%%%%%%%%%%%%%%%%%%%%%%%%%%%%%%%%%%%%%%%%%%%%%%%%%5

%%%%%%%%%%%%%%%%%%%%%%%%%%%%%%%%%%%%%%%%%%%%%%%%%%%%%%%%%%%%%%5

\subsection{Two bottlenecks: Edge effects}

Next we simulated a system with two bottlenecks where one is near the
boundary and one is far away. We concentrated on the case, 
where the bulk bottleneck is larger than that close to the boundary.

\begin{figure*}[ht]
\begin{center}
\vspace{0.5cm}
\includegraphics[height=5cm]{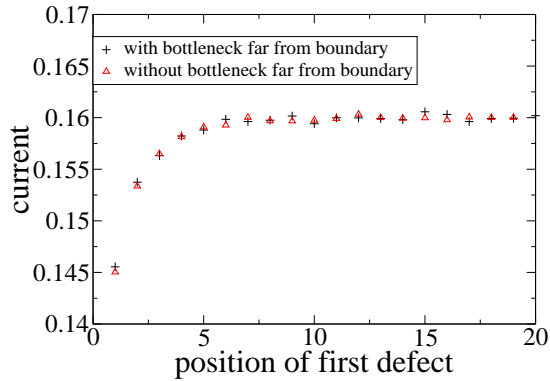}
\caption{\label{J(defpos)&bn}Dependence of the current on the position of 
  the first defect for $\alpha=0.2$ and $\beta=1.0$. Comparison of a
  system with only one defect and a system with a additional
  bottleneck far from the boundaries.}
\end{center}
\end{figure*}

In Fig.~\ref{J(defpos)&bn}, we plotted the dependence of the current
on the distance of the first bottleneck. For comparison we included
the results for a single bottleneck near the boundary from section 
\ref{one_bottleneck_at_boundary}.
One observes no significant difference between the two datasets. This
observation indicates that \emph{bottlenecks far from the boundary do
  not have any influence on the current}, as long as the current is
below the maximum current allowed by that bottleneck.

\begin{figure*}[ht]
\begin{center}
  \vspace{0.7cm}
\includegraphics[height=5cm]{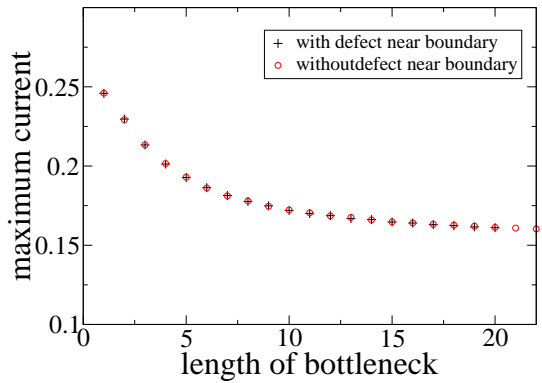}
\caption{\label{Jmax(l)&defect}Maximum current in dependence on the 
  bottleneck length for $q=0.6$. The addition of a defect near the
  boundaries does not alter the maximum current. }
\end{center}

\end{figure*}

In Fig.~\ref{Jmax(l)&defect}, however, we see the maximum current in a
system with a bottleneck of length $l$ far from the boundaries and a
defect at site 3. Again, one does not see a significant difference:
\emph{a small bottleneck near the boundary has no influence on the
  transport capacity}.

In agreement with observations already made in \cite{chou},
our results motivate the view of a local influence of
bottlenecks that yields the possibility to generalize concepts of the
TASEP with single bottleneck to systems with many bottlenecks. Only
bottlenecks near the boundaries have influence on the current if it is
below the transport capacity.  We therefore propose that the influence
of boundary defects can be 
taken into account through
effective boundary rates in the same manner as for a single bottleneck 
near the boundaries. That means that below the maximum current we can 
describe the system as a pure TASEP, but with effective boundary rates
$\alpha_{\rm eff},\beta_{\rm eff}$ depending on the configuration of
bottlenecks near the boundaries instead of the pure ones.

%%%%%%%%%%%%%%%%%%%%%%%%%%%%%%%%%%%%%%%%%%%%%%%%%%%%%%%%%%%%%%%%%%%%%%%%
%%%%%%%%%%%%%%%%%%%%%%%%%%%%%%%%%%%%%%%%%%%%%%%%%%%%%%%%%%%%%%%%%%%%%%%%
\section{Discussion and Outlook}

We have investigated the TASEP in the presence of a bottleneck of $l$
consecutive slow sites. Apart from computer simulations we have
developed an approximate analytical approach called \emph{interacting
  subsystem approximation (ISA)} that makes use of the exact results
obtained for the pure TASEP.  ISA turns out to be an efficient and
accurate method for systems with a single bottleneck.  It can be
applied to other systems as long as exact results are available.

Therefore we have divided the system into subsystems consisting of
fast and slow sites only, where the exact solution
\cite{derridaEHP,schuetz} for the homogeneous system applies.  In the
treatment of the coupling of these systems certain correlations are
neglected. Nevertheless the predictions of the analytical approach are
in good agreement with the simulations, while they slightly
underestimate the real current.  Our method yields much better results
than e.g.\ the mean field approach in \cite{kolomeiski2} for single
defects. For longer bottlenecks it is much more computationally
efficient than the approach by Chou and Lakatos \cite{chou}, which
produces results of a similar accuracy.  The {\em finite segment mean
  field theory (FSMFT)} introduced in \cite{chou} takes into account
only particle correlations within the bottleneck section and a certain
number of sites adjacent to it.  This approach is more flexible in the
kind of systems that can be treated, but it is algorithmically more
complex and less efficient since it requires the determination of the
leading eigenvector of a matrix that grows exponentially in the
segment length $l$.  As a consequence it can not be applied to systems
with longer bottlenecks $(l\geq 20$). In contrast, the ISA method
relies on finding \emph{one} root of a polynomial whose degree is
approximately $l$. This root can be isolated by physical constraints.
For this problem fast algorithms exist making it possible to compute
the current for very large bottlenecks.  Moreover the accuracy of the
ISA even increases for longer bottlenecks.  Since later \cite{inprep}
we want to use our results for disordered systems with finite density
of defects, a method that can also be applied to long bottlenecks is
required and the ISA appears to be a suitable procedure for this task.

The ISA can be applied for calculating the transport capacity (maximum
current) and the transition point to the phase separated phase of a
system with a bottleneck far from the boundaries, as well as the
current of a system with a bottleneck near the boundaries.  
If the bottleneck is far from the boundaries, the main effect on the 
current is a lowering of the maximum current. It is independent 
of the bottleneck position, but things change if the bottleneck is
located near one of the boundaries.  Here interactions with the
particle input or output become important (\emph{edge effect}, for single
defect sites see \cite{dong}).  This leads to a change of the current.
ISA reproduces the increase of the transport capacity if defects are
moved towards the boundary which was found by Dong et al.\ \cite{dong}).
Additional to this \emph{positive} edge
effect, ISA predicts that for lower entry rates the current is
\emph{decreased} by a bottleneck near a boundary compared to the
situation where it is deep in the bulk. This \emph{negative} edge
effect is confirmed
by numerical simulations. The simulations also show that there is a
regime where even a non-monotonic dependence on the bottleneck
position is possible. This effect, however, is \emph{not} reproduced by
ISA which indicates that it has its origin in correlations
that are neglected in the approximation.

A bottleneck near the boundary also has crucial effects on the phase
diagram. Indeed it leads to an \emph{absence} of a maximum
current-/phase separated phase!  One can say, the bottleneck near the
boundary softens the phase transition into a \emph{crossover}.  The
remaining phase transition line between low and high density phase,
on the other hand, is distorted compared to the pure system.
This observation motivates the concept of \emph{effective boundary
  rates}: the effects of the boundary bottleneck are
encompassed as effective boundary rates and the system can be
described as a pure one with the effective rates instead of the real
ones. One can calculate the relation between effective and real rates
by ISA which then can be used to determine the phase transition line.

In future work \cite{inprep} we will use the present results to derive
an effective theory for disordered systems which have a finite density
of defect sites even in the thermodynamic limit.  The results obtained
here for systems with two bottlenecks indicate that an accurate
description of the disordered case can be obtained, since the
``interaction'' of bottlenecks is rather local which is in agreement
with observations in \cite{chou} and \cite{dong}.  For large distances
of bottlenecks, the maximum current in the TASEP is dominated by the
longest bottleneck while (smaller) bottlenecks even close to the
boundaries do not have any influence on it. This observation supports
the conjecture made in \cite{barma2,krug} for periodic systems 
with quenched disorder that the longest bottleneck is dominating 
the transport capacity, and hence the transition to phase separation. 
Therefore it will be important to have an analytical
approach that can deal efficiently with very long bottlenecks.  In
contrast, if the system is not in the maximum current phase, we have
found that even large bottlenecks in the bulk do not have an influence
on the current.
These observations indicate that we can treat boundary effects and transport 
capacity independently for systems with many bottlenecks
by using the concept of effective boundary rates and the ISA
also in this case.

Natural applications of the results presented here are to biological
transport. The TASEP is the basis for the description of protein
synthesis and also active intracellular transport of motor proteins
along microtubuli. In both cases disorder plays an importat role through 
the inhomogeneity of the mRNA template \cite{slow_codon1,slow_codon2} 
or microtubuli associated proteins (MAPs) that attach to
microtubuli and can form obstacles for the motion of mobile motor
proteins like kinesin \cite{mandelkov,boehm}.

%%%%%%%%%%%%%%%%%%%%%%%%%%%%%%%%%%%%%%%%%%%%%%%%%%%%%%%%%%%%%%%%%%%%%%%%

\section*{\label{app_1}Appendix: More on edge effects}

We now want to determine a condition that the current decreases for a 
bottleneck approaching a boundary (\emph{negative edge effect}). 
Again, due to the particle-hole symmetry  (\ref{p-h-symmetry}) it is 
sufficient to consider a bottleneck near the left boundary.
We have to distinguish if the current for a defect at a small distance
$L_1={\mathcal O}(1)$ from the boundary is larger or less than the
current for a defect far from the boundary, $L_1\to\infty$. Though the
dependence of the current on the position of the first defect is not
always monotonic, for an approximation it is sufficient to compare the
current $J(L_1)$ for \emph{one} position of the defect near the boundary 
with the current $J(\infty)$ for a defect far from the boundary, since the 
mixed edge effect (see Sec.~\ref{one_bottleneck_at_boundary}) 
is quite weak and only in a small parameter region. For simplicity we 
take $L_1=2$.

Using (\ref{Z(N)}) and (\ref{Zp(N)}) yields
\begin{equation}
\label{edge-ineq1}
J(\alpha,\beta_1,2)=\frac{\alpha\beta_1^2 + \alpha^2\beta_1}{\beta_1^2 
+\alpha\beta_1 + \alpha^2 + \alpha\beta_1^2 + \alpha^2\beta_1} \, ,
\end{equation}
where $\beta_1$ is the virtual exit rate of subsystem 1. The edge
effect is negative if
\begin{equation}
\label{edge-ineq2}
J(\alpha,\beta_1,2)<J(\alpha,\beta_1,\infty) \, .
\end{equation}
In the L-phase\footnote{The term ``phase'' here refers to the phases
  of the system with a defect far from the boundaries where it is
  well defined.} we have $J(\alpha,\beta_1,\infty)=\alpha(1-\alpha)$
and (\ref{edge-ineq2}) is fullfilled if 
\begin{equation}
\label{neg_edge-eff-cond}
\beta_1<1-\alpha \, .
\end{equation}

Though we cannot determine $\beta_1$ exactly, we can state that
$\beta_1$ is less than in a pure system with $q=1$. However, in a pure
system there is a flat density profile without correlations, and we
have exactly $\beta_1=1-\alpha$, so (\ref{neg_edge-eff-cond}) is
always fulfilled. Therefore in the L-phase and, due to
particle-hole-symmetry, in the H-phase, the edge effect is always
negative.

In systems with many randomly distributed bottlenecks it is very
unlikely that longest one is near
the boundary. 
Thus the transition to the PS-phase, where the current is independent
of the boundary rates, will occur for lower rates of
$\alpha(\beta)$, where the negative edge effect is predominant.

%%%%%%%%%%%%%%%%%%%%%%%%%%%%%%%%%%%%%%%%%%%%%%%%%%%%%%%%%%%%%%%%%%%%%%%%
%%%%%%%%%%%%%%%%%%%%%%%%%%%%%%%%%%%%%%%%%%%%%%%%%%%%%%%%%%%%%%%%%%%%%%%%

\vspace{0.5cm}

\noindent{\bf Acknowledgements:}\\
We like to thank Joachim Krug for helpful discussions.

%%%%%%%%%%%%%%%%%%%%%%%%%%%%%%%%%%%%%%%%%%%%%%%%%%%%%%%%%%%%%%%%%%%%%%%%
%%%%%%%%%%%%%%%%%%%%%%%%%%%%%%%%%%%%%%%%%%%%%%%%%%%%%%%%%%%%%%%%%%%%%%%%


\begin{thebibliography}{99}

\bibitem{schmzia}
B. Schmittmann and R.P.K.\ Zia: 
%{\em Statistical Mechanics of Driven Diffusive Systems}, 
in C.\ Domb and J.L.\ Lebowitz (eds.),
{\em Phase Transitions and Critical Phenomena}, Vol.~17
(Academic Press, 1995)

\bibitem{derrida}
B. Derrida: Phys.\ Rep.\ {\bf 301}, 65 (1998)

\bibitem{gunterrev} G.M. Sch\"utz: {\em Exactly Solvable Models for
Many-Body Systems}, in C.\ Domb and J.L.\ Lebowitz (eds.),
{\em Phase Transitions and Critical Phenomena}, Vol.~19
(Academic Press, 2001)

\bibitem{css}
D.\ Chowdhury, L.\ Santen, and A.\ Schadschneider:
Phys.\ Rep.\ {\bf 329}, 199 (2000)

\bibitem{granflow}
H. Hayakawa and K. Nakanishi:
%{\em Universal behaviors in granular flows and traffic flows};
Prog. Theor. Phys. Suppl. {\bf 130}, 57 (1998)

\bibitem{frey1}
A. Parmeggiani, T. Franosch and E. Frey:
%\emph{Phase Coexistence in Driven One Dimensional Transport};
Phys. Rev. Lett. {\bf 90}, 086601 (2003)

\bibitem{lipowsky} 
R. Lipowsky, S. Klumpp, and T. M. Nieuwenhuizen:
%\emph{Random walks of cytoskeletal motors in open and closed compartments};
Phys. Rev. Lett. {\bf 87}, 108101 (2001)

\bibitem{NOSC1}
K. Nishinari, Y. Okada, A. Schadschneider, D. Chowdhury:
%{\em Intra-cellular transport of single-headed molecular motors KIF1A};
Phys. Rev. Lett. {\bf 95}, 118101 (2005)

\bibitem{NOSC2}
P. Greulich, A. Garai, K. Nishinari, A. Schadschneider, D. Chowdhury:
%{\em Intra-cellular transport by single-headed kinesin KIF1A: 
%effects of single-motor mechano-chemistry and steric interactions};
Phys. Rev.~{\bf E75}, 041905 (2007)

\bibitem{mcdonald}
C. MacDonald, J.Gibbs, A. Pipkin: 
%\emph{Kinetics of biopolymerization on nucleic acid templates};
Biopolymers {\bf 6}, 1 (1968) 

\bibitem{mpa}
R. Blythe, M.R. Evans:
%\emph{Nonequilibrium steady states of matrix product form: a solver's guide};
to appear in J. Phys. A (e-print arXiv:0706.1678)

\bibitem{krug1}
J. Krug:
%\emph{Boundary-induced phase transitions in driven diffusive systems}; 
Phys. Rev. Lett. 67, 1882 (1991) 

\bibitem{derridaEHP}
B. Derrida, M.R. Evans, V. Hakim, V. Pasquier:
%\emph{Exact solution of a 1D asymmetric exclusion model using a matrix 
%formulation};
J. Phys A {\bf 26} 1493-1517 (1993) 

\bibitem{schuetz}
G.M. Sch\"utz, E. Domany:
%\emph{Phase Transitions in an Exactly Soluble One-Dimensional Exclusion 
%Process}; 
J. Stat. Phys. {\bf 72}, 277 (1993)
 
\bibitem{koloSKS} 
A.B.\ Kolomeisky, G.~Sch\"utz, E.B.\ Kolomeisky and
J.P.\ Straley: J.\ Phys.\ A {\bf 31}, 6911 (1998)

\bibitem{Stinch02}
R.B. Stinchcombe:
%{\em Disorder in non-equilibrium models};
J. Phys. Cond. Matt. {\bf 14}, 1473 (2002)

\bibitem{Barma06}
M. Barma:
%{\em Driven diffusive systems with disorder};
Physica {\bf A372}, 22 (2006)

\bibitem{lebowitz}
S.A. Janowsky, J.L. Lebowitz:
%\emph{Finite-size effects and shock fluctuations in the asymmetric 
%simple-exclusion process}; 
Phys. Rev. A {\bf 45}, 618 (1992)

\bibitem{lebowitz2}
S.A. Janowsky, J.L. Lebowitz:
%\emph{Exact Results for the Asymmetric Simple Exclusion Process 
%with a Blockage}, 
J. Stat. Phys. {\bf 77}, 35 (1994) 

\bibitem{kolomeiski2}
A.B. Kolomeisky:
%\emph{Asymmetric simple exclusion model with local inhomogenity}; 
J. Phys. A {\bf 31}, 1153 (1998) 

\bibitem{chou}
T.\ Chou, G.W.\ Lakatos: 
%\emph{Clustered bottlenecks in mRNA translation and protein synthesis};
Phys. Rev. Lett. {\bf 93}, 198101 (2004)

\bibitem{dong}
J.J. Dong, B. Schmittmann, R.K.P. Zia: 
%\emph{Towards a model for protein production rates};
J. Stat. Phys. {\bf 128}, 21 (2007)

\bibitem{frey2}
P. Pierobon, M. Mobilia, R. Kouyos, E. Frey: 
%\emph{Bottleneck-induced transitions in a minimal model for intracellular 
%transport}; 
Phys. Rev. E {\bf 74}, 031906 (2006) 

%\bibitem{saad}
%Y. Saad,
%\emph{Numerical Solution of Large Eigenvalue Problems}
%(Halsted Press, New York, 1992).

\bibitem{inprep}
P. Greulich, A. Schadschneider: in preparation

\bibitem{barma}
G. Tripathy, M. Barma:
%\emph{Steady State and Dynamics of Driven Diffusive Systems with Quenched 
%Disorder}; 
Phys. Rev. Lett. {\bf 78}, 3039 (1997) 

\bibitem{barma2}
G. Tripathy, M. Barma:
%\emph{Driven lattice gases with quenched disorder: Exact results and 
%different microscopic regimes}; 
Phys. Rev. E {\bf 58}, 1911 (1997) 

\bibitem{krug}
J. Krug: 
%\emph{Phase Separation in Disordered Exclusion Models}; 
Braz. J. Phys. {\bf 30}, 97 (2000) 

\bibitem{harris1}
R.J. Harris, R.B. Stinchcombe:
%\emph{Disordered asymmetric simple exclusion process: Mean-field treatment}; 
Phys. Rev. E {\bf 70}, 016108 (2004) 

\bibitem{kolw}
K.M. Kolwankar, A. Punnoose: 
%\emph{Disordered totally asymmetric simple exclusion process: Exact results}; 
Phys. Rev. E {\bf 61}, 2453 (2000)

\bibitem{derrida2}
C. Enaud, B. Derrida:
%\emph{Sample-dependent phase transitions in disordered exclusion models}; 
Europhys. Lett. {\bf 66}, 83 (2004) 

\bibitem{lakatos}
G.W.\ Lakatos, T.\ Chou, A. Kolomeisky: 
%\emph{Steady-state properties of a totally asymmetric exclusion 
%process with periodic structure};
Phys. Rev. E {\bf 71}, 011103 (2005) 

\bibitem{juhasz}
R. Juhasz, L. Santen, F. Igloi: 
%\emph{Partially asymmetric exclusion processes with sitewise disorder};
Phys. Rev. E {\bf 74}, 061101 (2006) 

\bibitem{foulaad}
M.E. Foulaadvand, S. Chaaboki, M. Saalehi:
%\emph{Characteristics of the asymmetric simple exclusion process in the 
%presence of quenched spatial disorder}; 
Phys. Rev. E {\bf 75}, 011127 (2007)

\bibitem{slow_codon1}
M. Robinson \emph{et al.},
%\emph{Codon usage can affect efficiency of translation of genes In Escherichi coli},
Nucleic Acids Research, {\bf 12}, 6663 (1984)

\bibitem{slow_codon2}
M.A. Soerensen, C.G. Kurland and S. Pedersen:
%\emph{Codon Usage Determines Translation Rate in Escherichia coli};
J. Mol. Biol., {\bf 207}, 365-377 (1989)

\bibitem{mandelkov} 
E.-M. Mandelkow, K. Stamer, R. Vogel, E. Thies and E. Mandelkow:
%\emph{Clogging of axons by tau, inhibition of axonal traffic and 
%starvation of synapses},
Neurobiology of Aging, {\bf 24}, 1079-1085 (2003)

\bibitem{boehm} 
K.J. B\"ohm: private communication

\end{thebibliography}
\end{document}